# THE FREQUENCY FUNCTION OF ELLIPTICAL GALAXY INTRINSIC SHAPES

BENOIT TREMBLAY AND DAVID MERRITT
DEPARTMENT OF PHYSICS AND ASTRONOMY
RUTGERS UNIVERSITY, PISCATAWAY, NJ 08855

## ABSTRACT

We present fully nonparametric estimates of the frequency function of elliptical galaxy intrinsic shapes under the axisymmetric and triaxial hypotheses. If elliptical galaxies are assumed to be oblate or prolate, the frequency function of intrinsic shapes is negative for axis ratios near unity due to the lack of apparently round galaxies. Both axisymmetric hypotheses are found to be inconsistent at the 99% level with the data. Triaxial intrinsic shapes are fully consistent with the data; a number of possible triaxial frequency functions are presented, some of which exhibit strong bimodality. We also compute the "maximum entropy" distribution of intrinsic shapes under the triaxial hypothesis.

## 1. INTRODUCTION

Information about the three-dimensional shape of a galaxy is lost when the galaxy is projected onto the plane of the sky. This loss of information is acute in the case of elliptical galaxies, whose apparent shapes are elliptical but whose intrinsic shapes could be oblate, prolate or fully triaxial.

Beginning with Hubble (1926), many attempts have been made to deduce the distribution of intrinsic shapes of elliptical galaxies from the distribution of their apparent shapes, under the assumption of random orientations. This inverse problem has a formally unique solution if galaxies belong to a one-parameter family of shapes, e.g. oblate or prolate, but is degenerate if galaxies are triaxial. Thus a statistical approach is limited in what it can tell us about elliptical galaxy intrinsic shapes.

More recently, the emphasis has been directed toward increasing the quality of the data and the uniformity of the samples. Binney & de Vaucouleurs (1981) analyzed the ellipticities of 1750 galaxies in the *Second Reference Catalogue*. They found satisfactory agreement with the oblate, prolate and triaxial hypotheses. Ryden (1992) used CCD surface photometry of a sample of 171 bright ellipticals from Djorgovski (1985) to derive accurate, luminosity-weighted axis ratios. She then explored the degree to which these data could be described by a family of Gaussian frequency functions for the intrinsic shapes. In her best-fit model, the mode of the distribution lay near oblate galaxies with axis ratios of $\sim 0.7$. However distributions in which the most probable shape was nearly prolate were also permitted.

Here we reanalyze the Djorgovski-Ryden data set, but with a very different set of tools. We begin by noting that the questions we are addressing are "ill-conditioned" in the sense defined by statisticians (O'Sullivan 1986). An ill-conditioned problem is one for which a classical least-squares or maximum likelihood solution is not uniquely defined, or for which the sensitivity of such solutions to slight perturbations in the data is large. The estimation of the frequency function of galaxy intrinsic shapes is ill-conditioned in three distinct ways. First, the data are discrete, whereas we desire smooth continuous estimates of the unknown functions. This means that the data will have to be smoothed, and the form of the solution might depend strongly on the amount or character of the smoothing. Second, the distribution of intrinsic shapes is related to the distribution of apparent shapes via a deconvolution. Deconvolutions have the inherent property of amplifying finite-sample fluctuations in the data. Third, as mentioned above, even a completely specified distribution of apparent axis ratios is consistent with a wide range of different distributions over the *two* axis ratios that define a triaxial figure. Thus the problem is mathematically underdetermined unless we assume axial symmetry, or some other relation between two of the axis lengths.

A common way of dealing with ill-conditioned problems is to assume at the outset some parametric form for the solution, and then to vary the parameters until the fit to the data is optimized. Such an approach was adopted by Ryden (1992). While useful, Ryden's analysis has certain formal and practical disadvantages. Since the triaxial inverse problem is mathematically underdetermined, even with perfect data, Ryden's best-fit model must be an artifact of her assumed functional form. Very different functional forms could undoubtedly be found that fit the data equally well. Even assuming axisymmetry, parametric methods are biased in the sense that the assumed functional form will never be exactly correct; this misspecification may appear tiny when comparing the model to the data, but still – because of the ill conditioning – imply a large error in the function of most interest, the distribution of intrinsic shapes.

Another common approach to the solution of ill-conditioned problems is to use an interative technique like the Richardson (1972) - Lucy (1974) algorithm ("Lucy's method") to find a smooth solution that is consistent with the data. Lucy's method has been applied a number of times to the intrinsic shape frequency function problem (e.g. Binney & de Vaucouleurs 1981; Franx, Illingworth & de Zeeuw 1991; Fasano & Vio 1991). However Lucy's method enforces positivity of solutions, and we demonstrate below that the most likely function describing the distribution of intrinsic shapes under the axisymmetric hypothesis is significantly *negative* near axis ratios of unity.

In the present paper, we use modern function estimation techniques to construct a smooth estimate of $f(q)$, the distribution of apparent axis ratios, from the Djorgovski-Ryden sample. We then operate mathematically on this function to obtain estimates of the frequency function of intrinsic shapes. This procedure is "consistent" in the sense defined by statisticians, i.e. it gives exactly correct estimates in the limit of large sample sizes. The amplification of errors that results from the deconvolution is handled by making an appropriate choice of smoothing length when constructing $f(q)$.

We confirm the conjecture of Merritt (1990, 1992) and Fasano & Vio (1991) that the distribution of Hubble types is significantly inconsistent with either the oblate or prolate hypotheses due to the small number of nearly round galaxies. We then show, as did Ryden (1992) and others, that a distribution of triaxial intrinsic shapes can be fully consistent with the data. We present a number of such distributions obtained by assuming that all ellipticals are triaxial to the "same" degree. Many of these possible solutions are strongly bimodal. Finally, we derive the "maximum entropy" distribution of intrinsic shapes under the triaxial hypothesis.

## 2. THE AXISYMMETRIC INVERSE PROBLEM

Let the frequency function of apparent axis ratios be $f(q)$, $0 \leq q \leq 1$. Under the axisymmetric hypothesis, the frequency function of intrinsic axis ratios is also a one-dimensional function; call it $N(\beta)$, $0 \leq \beta \leq 1$. The relation between $f$ and $N$ under the oblate hypothesis, assuming random orientations, is

$$f(q) = q \int_0^q \frac{N_o(\beta)d\beta}{\sqrt{1-\beta^2}\sqrt{q^2-\beta^2}} \qquad (1)$$

(e.g. Mihalas & Binney 1981, p. 337) with inverse

$$N_o(\beta) = \frac{4}{\pi}\beta\sqrt{1-\beta^2}\frac{d}{d\beta^2}\int_0^\beta \frac{f(q)dq}{\sqrt{\beta^2-q^2}} \quad (2)$$
$$\equiv L_o f.$$

Under the prolate hypothesis, we have

$$f(q) = q^{-2}\int_0^q \frac{N_p(\beta)\beta^2 d\beta}{\sqrt{1-\beta^2}\sqrt{q^2-\beta^2}} \quad (3)$$

with inverse

$$N_p(\beta) = \frac{4}{\pi}\beta^{-1}\sqrt{1-\beta^2}\frac{d}{d\beta^2}\int_0^\beta \frac{f(q)q^3 dq}{\sqrt{\beta^2-q^2}} \quad (4)$$
$$\equiv L_p f.$$

Both $L_o$ and $L_p$ are linear operators. In a formal mathematical sense, the inverse problem is fully defined once the assumption of oblate or prolate symmetry is made.

The corresponding *statistical* problem is less well defined. Given a random sample of apparent axis ratios $Q_1, ..., Q_n$ drawn from the (unknown) function $f(q)$, what are the most likely distributions of true axis ratios $N_o(\beta)$ and $N_p(\beta)$?

Estimators for the functions $N_o$ and $N_p$ may be defined as follows. Construct a smooth non-parametric estimate of $f$ from the data; call this estimate $\hat{f}$. The estimate of $N_o$ is then defined as

$$\hat{N}_o = L_o \hat{f} \quad (5)$$

with a similar definition for $\hat{N}_p$. Based on theorems of Wahba (1990, p. 19), Scott (1992, p. 132) and others, such estimates are expected to be "consistent" in the sense of converging to the true functions $N$ in the limit that the sample size used to estimate $f$ is large, as long as the estimate of $f$ is itself asymptotically unbiased. All of the nonparametric estimators that we will use to construct $\hat{f}$ have this property. By contrast, parametric estimators are essentially guaranteed not to be consistent, since the assumed functional form is never exactly the same as the one that occurs in nature.

The major difficulties in applying this prescription when the sample size is not asymptotically large lie in the choice of nonparametric algorithm for representing $f$, and in the degree of smoothing to be applied to the data. Because the operators $L_o$ and $L_p$ are essentially differentiations, we expect the smoothness of any estimate $\hat{N}$ to be less than that of the corresponding $\hat{f}$. One result (e.g. Scott 1992, Theorem 6.19) is that the optimal degree of smoothing to be applied to the data when estimating $\hat{N}$ will generally be larger than if $\hat{f}$ itself were the function of interest. Choosing the appropriate degree of smoothing for each function in an objective way is a difficult problem, which we return to below.

### 2.1. *Maximum Penalized Likelihood Estimators*

Our first step is to construct a nonparametric estimate of $f(q)$. In the technique of "maximum penalized likelihood" function estimation, one searches for the function $\hat{f}$ that maximizes the functional

$$\log \mathcal{L}_p = \sum_{i=1}^n \log f(Q_i) - \lambda P(f) \quad (6a)$$

subject to the constraints

$$\int f(q)dq = 1, \quad f(q) \geq 0 \quad (6b)$$

(Thompson & Tapia 1990, p. 102). The first term in (6a) is the usual log likelihood; here, $f$ is assumed to be so general in its definition that virtually any functional form can be represented. The positive number $\lambda$ is the smoothing parameter and $P(f)$ is the penalty function. Maximizing (6) with $\lambda = 0$ would produce an estimate that is unacceptably noisy, consisting of spikes at most or all of the data points. The penalty function ensures that the solution will be more-or-less smooth, by assigning a penalty to solutions that are rapidly varying; increasing $\lambda$ increases the degree of smoothness, but also reduces the fidelity to the data, i.e. biases the solution.

The penalty function (really a functional, since it depends on the function $f$) is often chosen so that the solution remains physically reasonable even in the limit of large $\lambda$. A number of experiments with the Djorgovski-Ryden data set suggested that the function $f(q)$ is crudely describable as a Gaussian (see also Figure 6 of Ryden 1992). A natural choice for the penalty function is therefore the one suggested by Silverman (1982):

$$P(f) = \int_0^1 \left[(d/dq)^3 \log f\right]^2 dq. \tag{7}$$

Silverman's penalty function is a "Gaussian filter" that assigns zero penalty to any Gaussian $f$. As $\lambda$ increases, the limiting estimate obtained as the solution to the optimization problem (6) becomes the normal density with the same mean and variance as the data (Silverman 1982, Theorem 2.1). For smaller $\lambda$ the solution will reflect more closely the true density implied by the data, and $P(f)$ simply plays the role of a smoothing operator. By choosing $P(f)$ in this way, we guarantee that even a gross overestimate of the smoothing parameter $\lambda$ will yield, at worst, a Gaussian fit to the data. The kernel estimates discussed below do not have this property.

Following Silverman (1982), the solution $\hat{f}$ to the constrained optimization problem of equations (6) may be found more simply as the *un*constrained maximum of the functional

$$\sum_{i=1}^n \log f(Q_i) - \lambda P(f) - n \int_0^1 f(q) dq. \tag{8}$$

We obtained approximate solutions to this unconstrained optimization problem as follows. Define a uniform grid in $q$, $q_1, ..., q_m$, where $q_j = q_1 + (j-1)\delta$ and $\delta = (q_m - q_1)/(m-1)$. A discrete representation of (8) on this grid is

$$\sum_{i=1}^n \log f(Q_i) - \lambda \delta^{-5} \sum_{j=2}^{m-2} \left[-\log f_{j-1} + 3\log f_j - 3\log f_{j+1} + \log f_{j+2}\right]^2 - n \sum_{j=1}^m \epsilon_j f_j, \tag{9}$$

with $\epsilon_1 = \epsilon_m = \delta/2$, $\epsilon_j = \delta, j = 2, ..., m-1$. In the first term of equation (9), $f(Q_i)$ is computed by assuming a linear dependence of $f$ on $q$ between the grid points $k$ and $l$ such that $q_k < Q_i < q_l$.

One then simply varies the $m$ parameters $f_j$ in equation (9) until an approximate maximum is obtained. We chose $m = 50$, with $q_1 = 0.4$ and $q_m = 1$, and used the NAG routine E04JAF to carry out the optimization; a typical calculation required $\sim 30$ seconds on a Sparcstation 10.

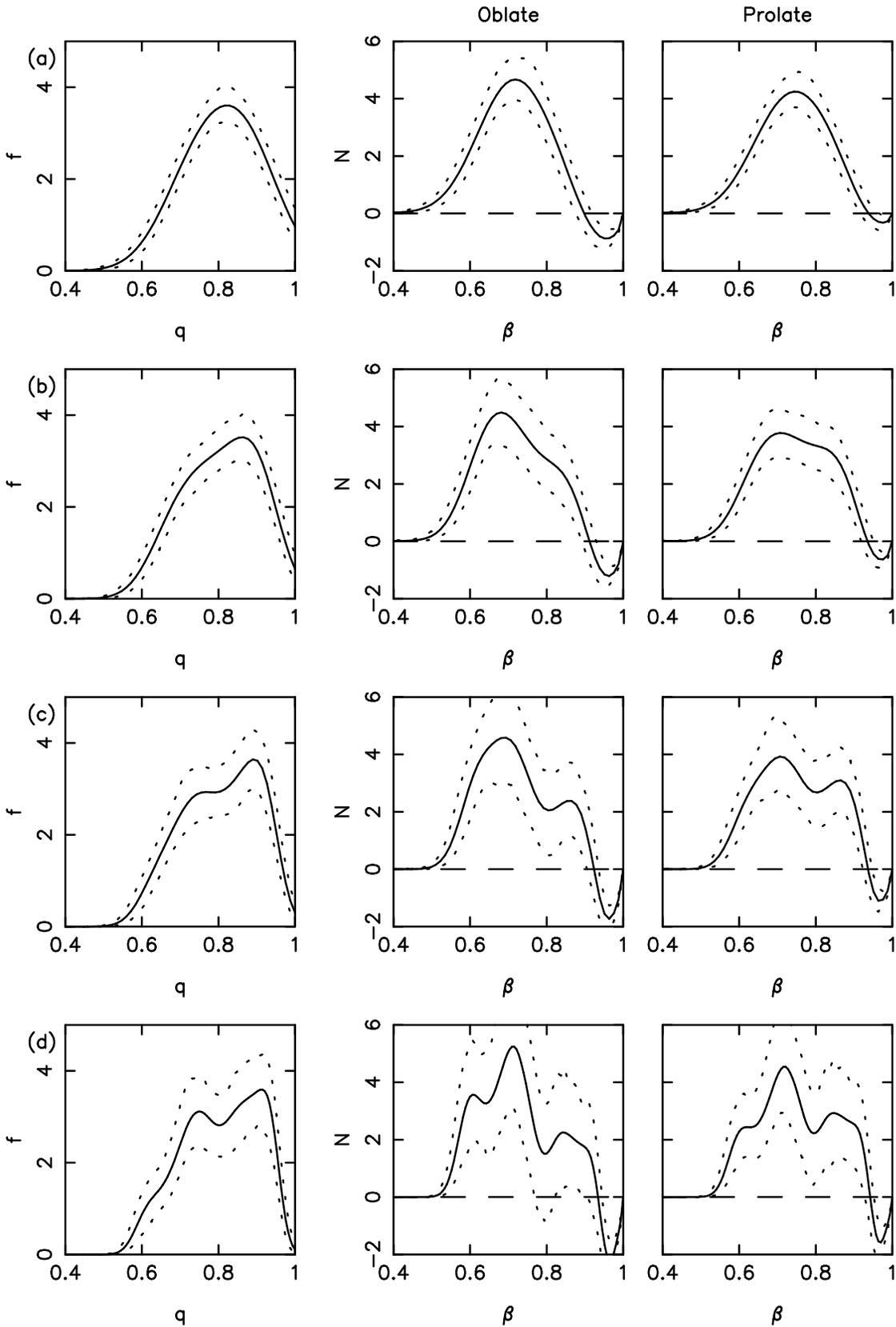

Figure 1. Maximum penalized likelihood estimates of $f(q)$, $N_o(\beta)$ and $N_p(\beta)$ made from the Djorgovski-Ryden sample. Dashed lines are 90% confidence intervals on the estimates. (a) $\lambda = 10^{-7}$; (b) $\lambda = 10^{-8}$; (c) $\lambda = 10^{-9}$; (d) $\lambda = 10^{-10}$.

Figure 1 shows estimates of $f$ and $N$ for four choices of the smoothing parameter $\lambda$. Approximate, 90% confidence bands on the estimates were obtained by bootstrap resampling from the data (Efron 1982); 200 bootstrap resamples were used. These confidence bands represent the expected point-by-point variance in the estimates due to the finite size of the data set (Scott 1992, p. 259).

For large $\lambda$, the estimates $\hat{f}_\lambda(q)$ are essentially Gaussian as expected; the mode lies near $q = 0.81$. As $\lambda$ is reduced, $\hat{f}_\lambda$ begins to exhibit bimodal structure, with modes at $q \approx 0.90$ and $q \approx 0.76$. Reducing $\lambda$ still further produces additional peaks that no doubt reflect sampling noise in the data.

For every choice of smoothing parameter $\lambda$, the estimates $\hat{f}_\lambda$ fall off sharply for $q \gtrsim 0.9$. Inspection of the curves produced with small $\lambda$ suggests that the true function $f$ is zero in the neighborhood of $q = 1$, i.e. that there are essentially no apparently round galaxies. The non-zero value of $f_\lambda(q = 1)$ for larger $\lambda$ is arguably just due to the smoothing effect of $P(f)$.

The ill-conditioning of the deconvolution operator is apparent in Figure 1. Modest changes in the appearance of $\hat{f}$ correspond to much larger changes in $\hat{N}$.

It is of some interest to know which of these estimates are the best representations of the "true" frequency functions $f$ and $N$ (assuming, in the case of $N$, that galaxies are actually axisymmetric). One usually defines the error of an estimate $\hat{g}$ as the integrated square error, or

$$\text{ISE} = \int (\hat{g} - g)^2 = \int \hat{g}^2 - 2\int \hat{g}g + \int g^2, \tag{10}$$

with $g$ the (unknown) true frequency function. Our goal is to choose $\lambda$ so as to minimize the ISE, or some estimate of the ISE, of $\hat{f}$ or $\hat{N}$.

A large number of algorithms have been described for achieving this goal when $g$ is the frequency function represented by the data, or $f(q)$ in the present case (e.g. Scott 1992, ch. 6). Much less has been written about techniques for minimizing the ISE of a function that is related to the data via a deconvolution, such as $N(\beta)$. Unfortunately it is the latter, more difficult problem that is of most interest here.

We begin by considering the optimal choice of $\lambda$ for estimating $f(q)$. Setting $g = f$ in equation (10), we note that the last term does not depend on $\hat{f}$. We therefore seek a $\hat{f}$ that minimizes the sum of the first two terms. The difficulty with this prescription is that the true solution $f$, which appears in the second term, is unknown.

The idea of "cross-validation" (Stone 1974) is commonly used to estimate the second term in equation (10). The "unbiased cross-validation" estimate of $\lambda$ is the value that minimizes UCV($\lambda$), where

$$\text{UCV}(\lambda) = \int \hat{f}^2 dq - \frac{2}{n}\sum_{i=1}^{n} \hat{f}_{-i}(Q_i) \tag{11}$$

(e.g. Scott 1992, p. 166; Silverman [1986, p. 48] calls this the "least-squares cross-validation" estimate.) Here $\hat{f}_{-i}$ is an estimate of $f$ constructed by leaving out the single datum $Q_i$. Thus, both terms on the right hand side of equation (11) may be estimated directly from the data.

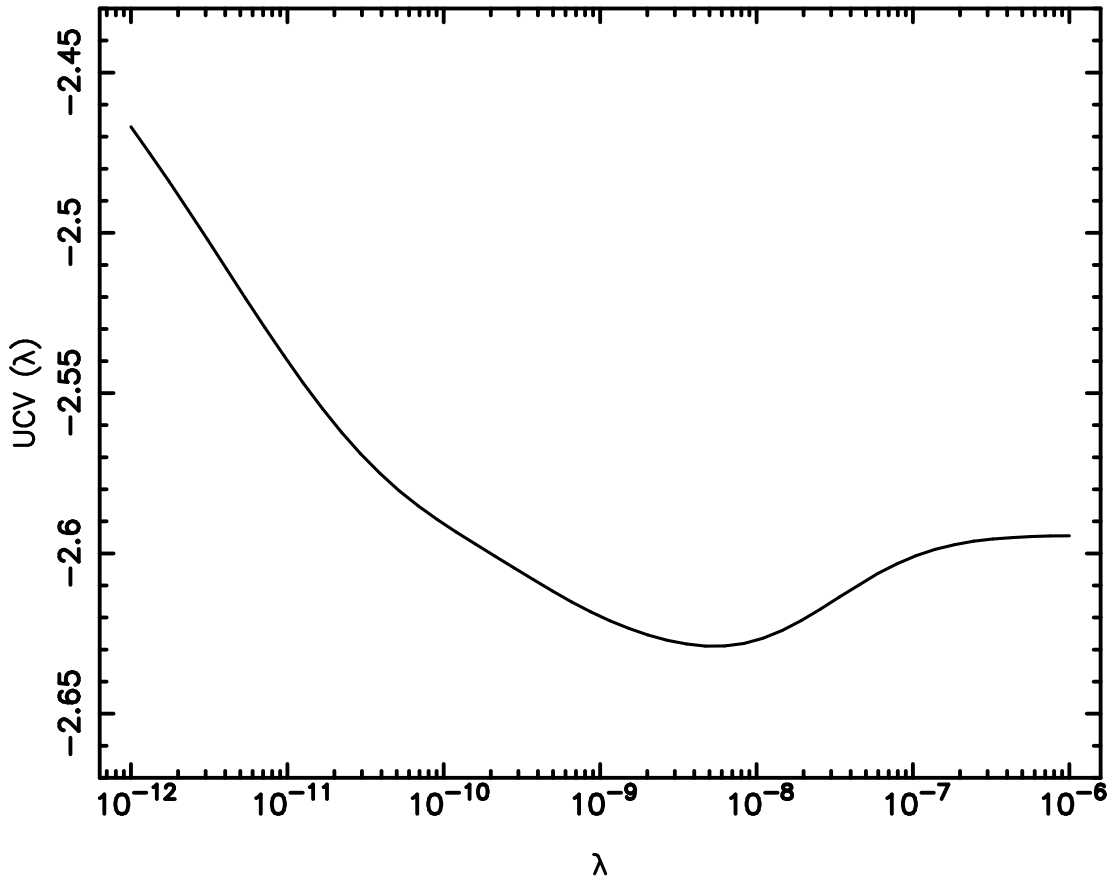

Figure 2. Cross validation estimate of the optimal smoothing parameter $\lambda$ for maximum penalized likelihood estimation of $f(q)$. The quantity UCV ($\lambda$) is equal, within an additive constant, to a cross-validation estimate of the integrated square error of $\hat{f}_\lambda(q)$ (see equation 11). The minimum lies at $\lambda \approx 5.5 \times 10^{-9}$.

Figure 2 shows the dependence of UCV on $\lambda$ for the Djorgovski-Ryden data set. There is a local minimum at $\hat{\lambda} \approx 5.5 \times 10^{-9}$. The corresponding, optimal estimate $\hat{f}_\lambda$ looks very similar to the second curve from the top in Figure 1 (obtained by setting $\lambda = 10^{-8}$): it exhibits a single mode, with a rapid falloff above $q \approx 0.9$.

As remarked above, the ideal $\lambda$ from the point of view of minimizing the ISE of $\hat{f}$ need not be equal to the ideal $\lambda$ for minimizing the ISE of $\hat{N}_o$ or $\hat{N}_p$. In fact we expect the optimal values of $\lambda$ for estimating the latter functions to be larger than the optimal value for estimating $f$, since $N$ is effectively a deconvolution or derivative of $f$, and the process of differentiation tends to increase the noise in the estimate (e.g. Scott 1992, p. 131). However a careful search of the statistical literature failed to turn up any very definite prescription for optimizing the smoothing in deconvolution problems like this one. Scott (1992, p. 265) remarks simply that "This class of problems [i.e. deconvolutions] is the most difficult in smoothing". Wahba (1990, p. 105) notes that the estimate of the optimal $\lambda$ in deconvolution problems is likely to itself be ill-conditioned, and recommends simply using the value obtained from minimizing UCV($\lambda$). Rice (1986) finds that "ordinary cross-validation and related techniques can be quite unsatisfactory" for choosing the optimal smoothing parameter in deconvolution problems.

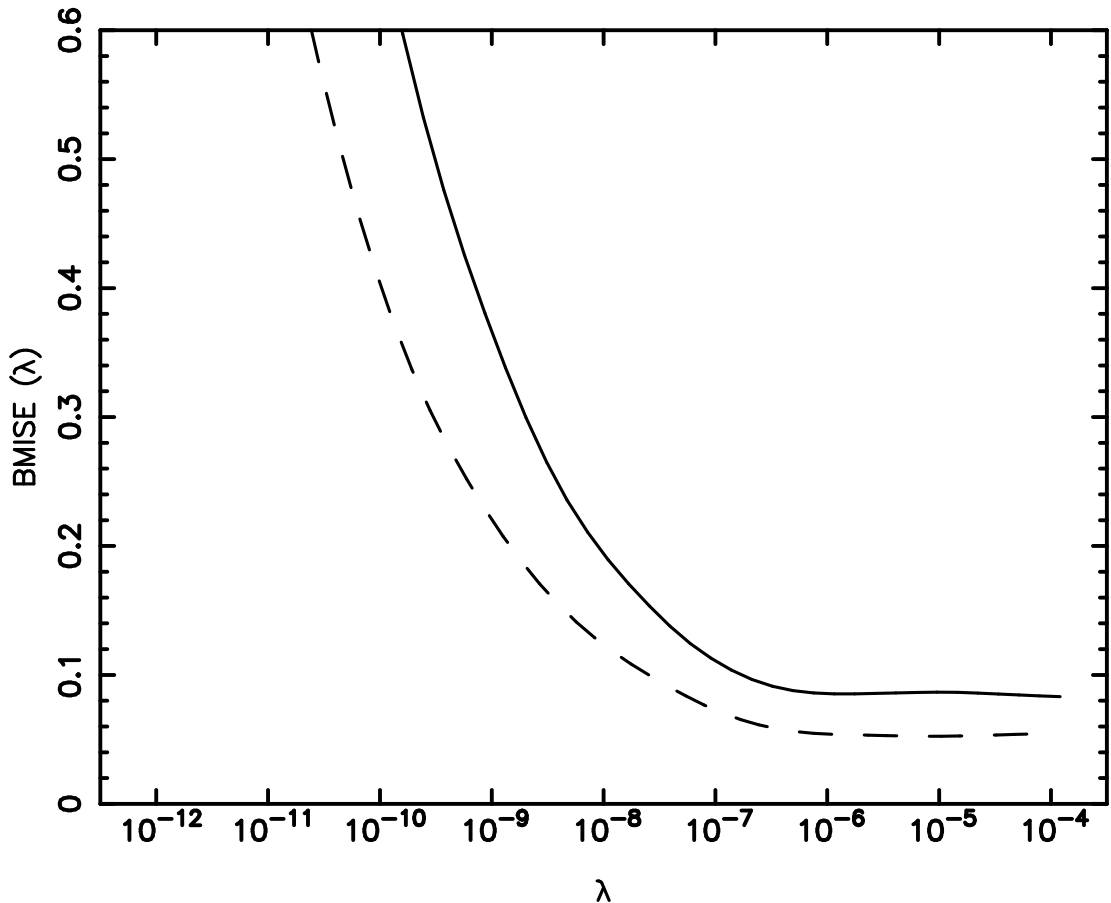

Figure 3. Bootstrap estimates of the mean integrated square error of maximum penalized likelihood estimates of $N_o(\beta)$ (dashed line) and $N_p(\beta)$ (solid line). Both curves show weak minima at large, oversmoothed values of $\lambda$.

We nevertheless made some attempts to find the optimal $\lambda$ for estimation of $N_o$ and $N_p$. Following the ideas of Taylor (1989), Hall (1990) and others, we defined a bootstrap estimate of the optimal $\lambda$ as the value that minimizes

$$\text{BMISE}(\lambda) = \int_0^1 [\hat{N}^*(\beta) - \hat{N}(\beta)]^2 d\beta. \qquad (12)$$

Here $\hat{N}$ is the estimate itself, constructed from the data with smoothing parameter $\lambda$. $\hat{N}^*$ is the estimate of $N$ using "data" $Q_i^*$ resampled, with replacement, from $\hat{f}_\lambda$. We stress that our criterion is intuitive and may be less than optimal. In any case, Figure 3 shows that BMISE $(\lambda)$ for both $N_o$ and $N_p$ have local minima only at very large, and almost certainly oversmoothed, values of $\lambda$. Such behavior is common when data sets are too small to usefully constrain the smoothing (e.g. Scott 1992, Fig. 6.17) and that is how we interpret our discouraging result.

We conclude that it is difficult or impossible to choose between the various estimates of $N_o$ or $N_p$ given in Figure 1, except that, as noted previously, the optimal estimates probably have values of $\lambda$ larger than the optimal value $\hat{\lambda} \approx 5.5 \times 10^{-9}$ found for the estimation of $f$. A reasonable choice of $\lambda$ for estimating $N_o$ or $N_p$ might be $\hat{\lambda} = 1.0 \times 10^{-8}$. We will adopt this value below when considering the triaxial inverse problem.

Although we can not say which of the estimates in Figure 1 for $N_o$ or $N_p$ are optimal, we note that all of our estimates have some features in common. In particular, for *every* choice of smoothing parameter $\lambda$ considered here, $N_o$ and $N_p$ exhibit dips below zero for axis ratios close to unity. The interval of negative $N$ is $0.90 \lesssim \beta < 1$ in the case of the oblate deconvolution, and $0.94 \lesssim \beta < 1$ in prolate case. Furthermore the 90% confidence bands suggest that these dips below zero are statistically significant, again for virtually every choice of $\lambda$: larger values of $\lambda$ produce less pronounced dips for $N(\beta)$, but the confidence bands also become tighter and the significance of the dip remains large.

For values of the smoothing parameter close to the "optimal" value $\hat{\lambda} = 5.5 \times 10^{-9}$, we find that the 99% confidence bands for both $\hat{N}_o$ and $\hat{N}_p$ (not shown in Figure 1) fall below zero for $\beta \approx 1$. Thus we conclude that the axisymmetric hypothesis can be rejected with confidence.

The reason for the negative values of $N$ is easy to find. A distribution of apparent axis ratios $f(q)$ is consistent with the oblate hypothesis only if the quantity

$$\int_0^q \frac{f(x)dx}{\sqrt{q^2 - x^2}} \tag{13}$$

is an increasing function of $q$; violation of this condition means that $N_o(\beta)$ will be negative for some values of $\beta$. Similarly, the prolate hypothesis is tenable only if the quantity

$$\int_0^q \frac{f(x)x^3 dx}{\sqrt{q^2 - x^2}} \tag{14}$$

is monotonically increasing. All of the estimates of $f$ shown in Figure 1 violate these conditions for $q$ near one, due to the rapid decline of $\hat{f}$ for $q \gtrsim 0.9$. This means that the number of nearly round galaxies in the Djorgovski-Ryden sample is too few to be reproduced by random orientations of axisymmetric spheroids, which have a tendency – larger for oblate than for prolate spheroids, but large for both – to appear nearly round in projection.

One possible interpretation of the negativity of $\hat{N}_o$ and $\hat{N}_p$ is that most of the galaxies in this sample are triaxial. As we demonstrate below, a distribution of triaxial intrinsic shapes can in fact naturally reproduce the rapid decline in $f$ for $q \approx 1$.

Another possibility is that most elliptical galaxies exhibit some small departures from ellipsoidal symmetry, which might cause otherwise nearly spherical galaxies to be classified as elongated. These departures could result from causes that are unrelated to the galaxies' true shapes – such as dust absorption, superimposed satellite galaxies, etc. – or could be intrinsic, such as imbedded disks, Malin-Carter "shells," tidal distortions, twists, etc. Certainly such features are known to be present in a large fraction of ellipticals (e.g. Schweizer & Seitzer 1992). The amplitude of the required effect is apparently $\sim 0.1$ in $q$, which is large compared to any expected measurement errors but not otherwise unreasonable.

We note that Ryden's (1992) algorithm for computing "average" axis ratios is particularly sensitive to such contamination, since she averages the axis ratio of each isophote without regard to its orientation. Any "noise" in the image will therefore tend to produce an increase in the average axis ratio as computed by her. One could imagine computing an average $q$ differently, by finding the single axis ratio of a *set* of ellipsoids, with fixed position angle, that best reproduces the luminosity distribution over some part of the image. Such an algorithm would be less influenced by distortions that are distributed randomly in position angle over the galaxy image.

Of course, the presence of such distorting features, if they are intrinsic, would by itself constitute evidence that many elliptical galaxies are not axisymmetric. Our point here is simply that a statistical invalidation of the axisymmetric hypothesis does not necessarily imply that the galaxies in our sample are all triaxial ellipsoids.

A number of authors have noted the difficulty of satisfying the axisymmetric hypothesis with data sets similar to ours (e.g. Merritt 1990, 1992; Fasano & Vio 1991). Here we have placed those earlier conclusions on a firmer statistical basis.

Our results emphasize the dangers of approaching the intrinsic shape frequency function problem using an algorithm like Richardson (1972) - Lucy (1974) iteration ("Lucy's method"). Lucy's method enforces positivity of the deconvolved function and is inappropriate for any problem where that function might want to dip below zero. (Of course, almost any deconvolution problem can fall into this category, not just the one considered here.) In addition, Lucy's method deals in a relatively inflexible way with the ill-conditioning of the deconvolution.

## 2.2. *Kernel Estimates*

The conclusions just reached might be contingent on the algorithm used to construct estimates of $f$. Accordingly, we explored a second, very different technique based on kernels for constructing nonparametric estimates of $f$. The use of kernels in this context was apparently first considered by Fasano & Vio (1991).

The kernel estimate of a frequency function $f$ is defined as

$$\hat{f}(x) = \frac{1}{nh} \sum_{i=1}^{n} K\left(\frac{x - X_i}{h}\right) \tag{15}$$

where $K(t)$, the "kernel function", satisfies

$$\int_{-\infty}^{\infty} K(t) dt = 1. \tag{16}$$

$X_i$ is the value of the $i^{th}$ measurement, and $h$ is the window width (also called the "bandwidth" or the "smoothing parameter".) The kernel estimate is a sum of "bumps" placed at the observations; the kernel function $K$ controls the shapes of the bumps while $h$ determines their widths. If $h$ is chosen appropriately, the kernels will overlap and the resultant estimate will be smooth and continuous.

The statistical literature gives some standard choices for the function $K(t)$ :

a) Quadratic, $K(t) = \frac{3}{4}(1 - t^2)$;

b) Quartic, $K(t) = \frac{15}{16}(1 - t^2)^2$;

c) Gaussian, $K(t) = \frac{1}{\sqrt{2\pi}} e^{\frac{-t^2}{2}}$.

Both the quadratic and the quartic kernels are supported on $-1 \leq t \leq 1$, while the Gaussian kernel has infinite support. All authors agree that the choice of a particular form for the function $K$ has little influence on the estimate, a conclusion that is confirmed by our experience. We chose to use the quartic kernel for this study.

When a set of data is strongly inhomogeneous, a single value of $h$ will yield an estimate that is too smooth in the high-density regions and too noisy in the low-density regions. It is common practice to vary the window width in inverse proportion to the square root of the local density (Abramson 1982); we follow that practice here. The "adaptative kernel estimate" is

$$\hat{f}(x) = \frac{1}{nh} \sum_{i=1}^{n} \lambda_i^{-1} K\left(\frac{x - X_i}{h\lambda_i}\right), \tag{17a}$$

(Silverman 1986, p. 100), where

$$\lambda_i = \left[\frac{\tilde{f}(X_i)}{g}\right]^{-\frac{1}{2}} \tag{17b}$$

The "pilot estimate" $\tilde{f}(X_i)$ is an estimate of the density obtained using a fixed window width $h$, and $g$ is the geometric average of the $\tilde{f}(X_i)$.

The distribution of observed axis ratios is defined on the bounded interval [0,1]. A kernel estimate as defined above has no built-in boundaries, and will extend beyond the physical limits unless $h$ is extremely small. In our case, only the boundary at $q = 1$ is important. The literature presents at least four methods for overcoming the boundary problem:

1. Ignore the boundary, in which case the estimate is no longer a proper density over the allowed interval and gives weight to values of $q > 1$. This option was rejected.

2. Use a "boundary kernel" whose shape depends on the location of the data point $X_i$ (e.g. Scott 1992, p. 146). Near the boundaries, the kernel becomes asymmetric to avoid "spilling over" the edges. This method has the disadvantage of requiring a tailor-made kernel for each situation, and Scott (1992, p. 148) remarks that "severe artifacts can be introduced" by making an inappropriate choice. Indeed, we found that most simple modifications of Scott's prescription for a boundary kernel led to patently unphysical behavior of the estimates near $q = 1$. We decided to reject this option as well.

3. Extend the sample symmetrically beyond the boundaries. This is the so-called "reflection method" (e.g. Silverman 1986, p. 31). The main disadvantage is that reflection forces the derivative of the estimate to be zero at the boundary (in our case, at $q = 1$). Nevertheless this method is simple and robust and worth exploring.

4. Use a change of variables that moves the boundary to infinity. Many choices for the transformation exist; after some experimenting, we found that a simple log-transform $q' = -\ln(1 - q)$ was satisfactory. Unlike the case of reflection, this transformation guarantees that the estimate will have zero value at the boundary $q = 1$, rather than zero derivative. This method also seemed to give satisfactory results.

Figures 4 and 5 show kernel estimates of $f(q)$ and $N(\beta)$ constructed via the reflection and transformed-variable methods, for four choices of smoothing parameter $h$ or $h'$ ($h'$ is the window width in the transformed space). The 90% bootstrap confidence bands are also shown. The estimates obtained under reflection look reasonably similar to those obtained via maximum penalized likelihood, except for the behavior near $q = 1$. When the window width is sufficiently large, $f$ is greatly inflated near $q = 1$ and the distribution of intrinsic shapes $N(\beta)$ becomes fully nonnegative; this occurs first with the prolate deconvolution.

The estimates constructed using the change of variable look systematically different from the maximum penalized likelihood estimates: they appear to be oversmoothed at small $q$ and undersmoothed at large $q$. Both of these effects result from carrying out the smoothing in a transformed space. However the deconvolved functions are always significantly negative in the region near $q = 1$.

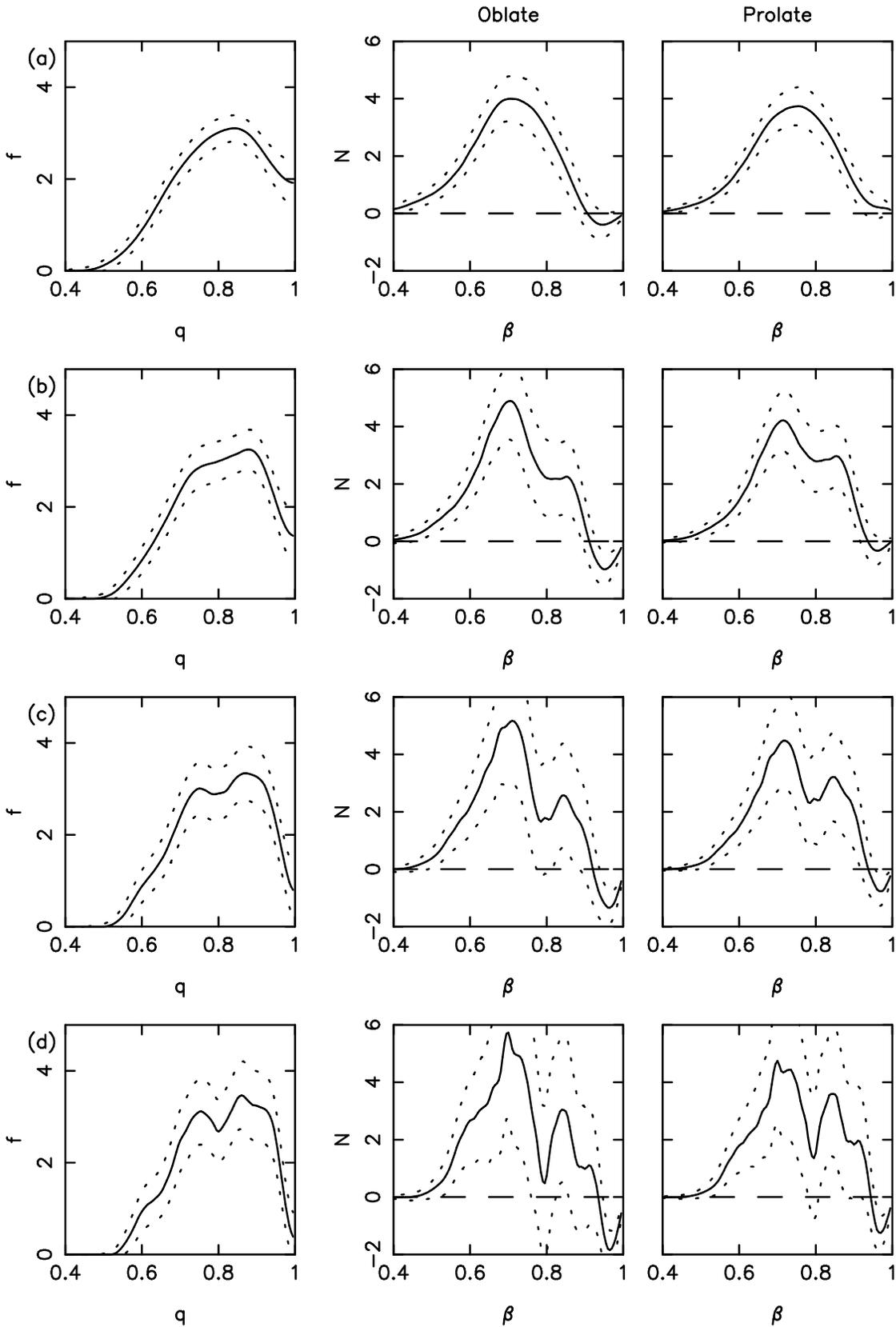

**Figure 4.** Estimates of $f(q)$, $N_o(\beta)$ and $N_p(\beta)$ constructed via the adaptive kernel algorithm, with reflection of the data around $q = 1$. Dashed lines are 90% confidence intervals on the estimates. (a) $h = 0.15$; (b) $h = 0.10$; (c) $h = 0.07$; (d) $h = 0.05$.

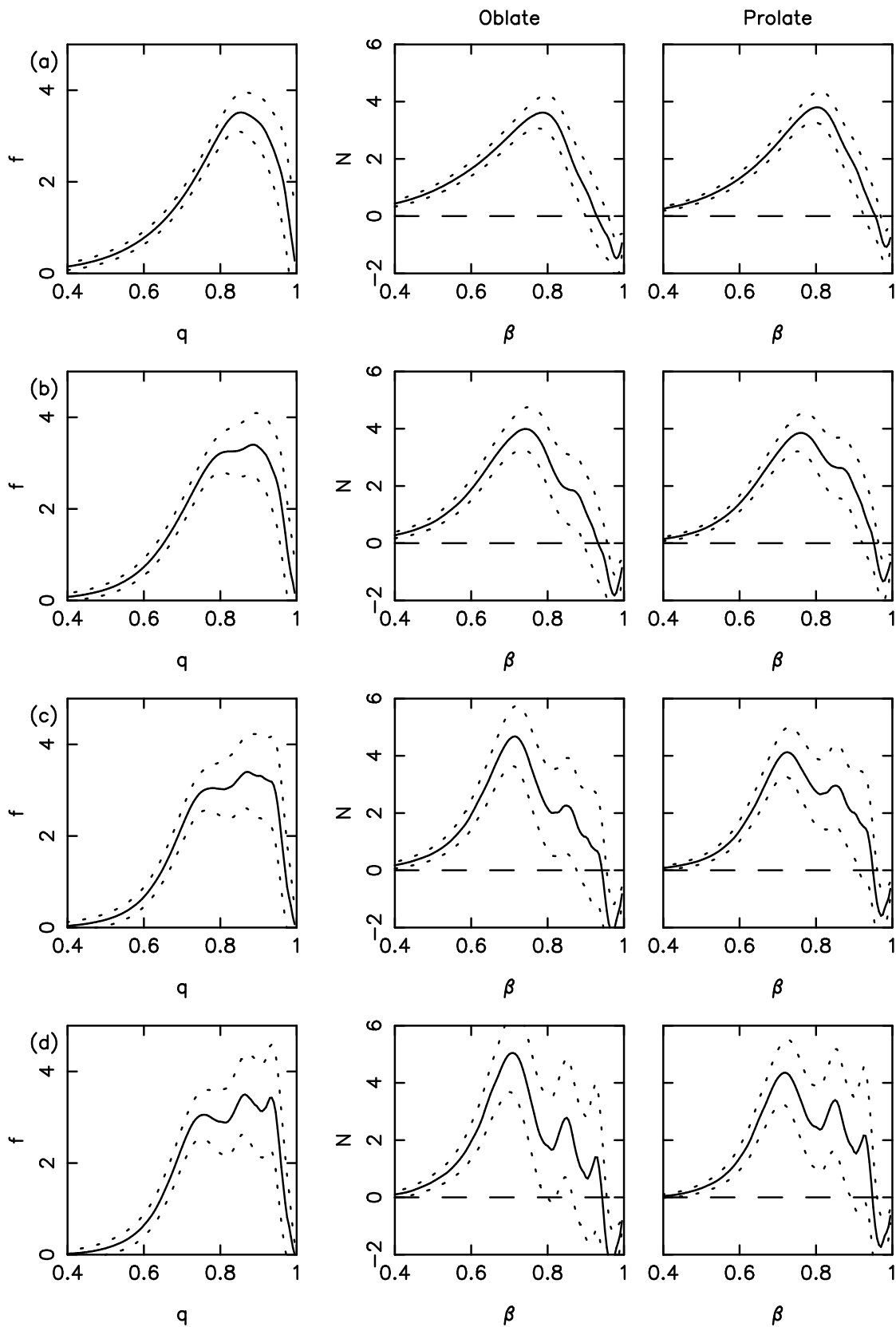

Figure 5. Like Fig. 4, except that estimates were constructed in the space of the transformed variable $q' = -\log(1-q)$. (a) $h' = 0.4$; (b) $h' = 0.5$; (c) $h' = 0.7$; (d) $h' = 1.0$.

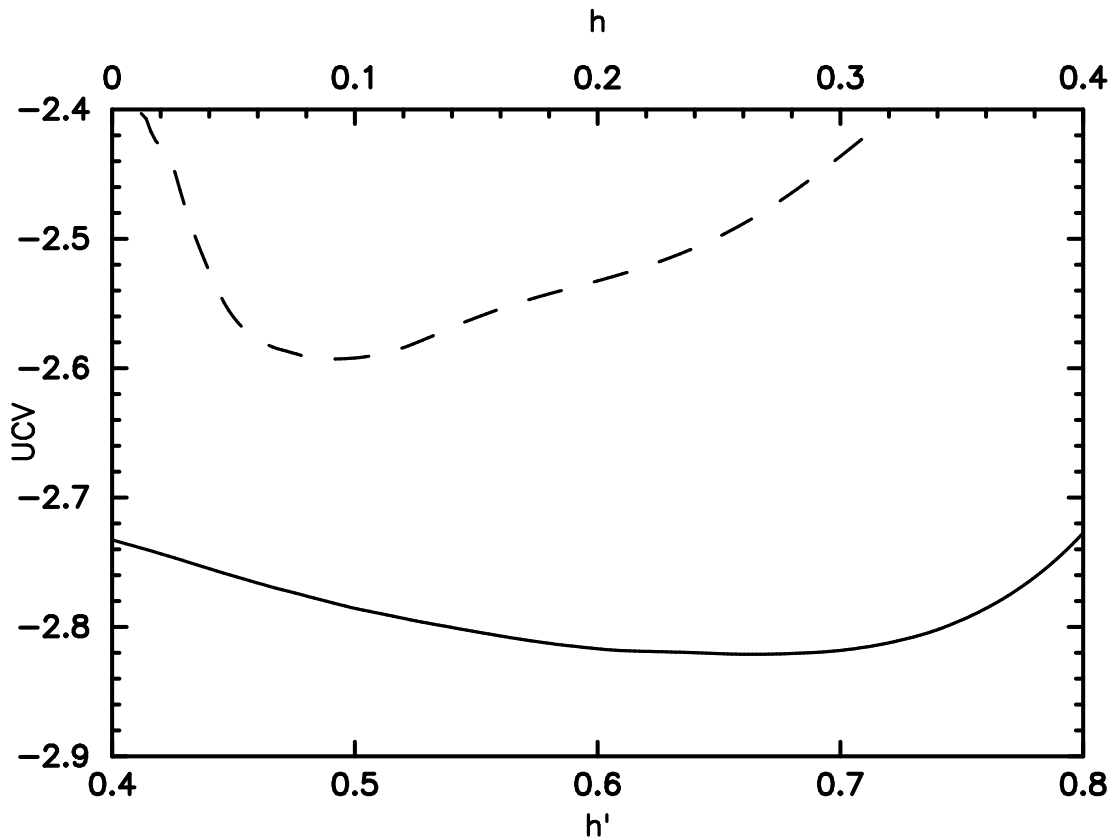

Figure 6. Cross validation estimates of the optimal smoothing parameter $h$ for kernel estimation of $f(q)$. Solid line: UCV($h'$); dashed line: UCV($h$).

The techniques discussed above for the choice of smoothing parameter in the case of a maximum penalized likelihood estimate can readily be applied to kernel estimates, and in fact these techniques were first discussed in the context of kernels (e.g. Scott 1992, ch. 6). Figure 6 shows UCV as a function of smoothing parameter for the two types of kernel estimate. The optimal values of the smoothing parameter are $h \approx 0.09$ and $h' \approx 0.65$.

The bootstrap algorithm described above for choosing the optimal value of the smoothing parameter in the construction of $N$ (Eq. 12) can be used here as well. Unfortunately, this method returned grossly oversmoothed estimates of $N$ as "optimal." Again, we conclude that our data set is too small to usefully constrain the optimal degree of smoothing in the derivation of $N_o$ or $N_p$.

### 2.3. *One Population or Two?*

It is intriguing that the "optimal" estimates of $f$, $N_o$ and $N_p$ using both the maximum penalized likelihood and kernel algorithms exhibit hints of bimodality. Here we investigate whether this bimodality is significant, and hence whether elliptical galaxies might be thought of as constituting two populations based on their apparent shapes. We ignore for the moment the fact that $N(\beta)$ is almost always significantly negative and therefore unphysical; our goal is simply to determine whether the degree of bimodality observed in the data is likely to be an artifact of the limited sample size.

Silverman (1981; 1986, p. 146) describes a nonparametric test for bimodality. (His test was designed for fixed kernel estimates but appears to generalize naturally to function estimates constructed in other ways.) Given a data set $X_i$, find the largest value of the smoothing parameter for which the estimate $\hat{f}(x)$ still exhibits bimodal structure (i.e. for which $f$ has two local maxima). Call this critical smoothing parameter $\alpha_{crit}$, and the corresponding estimate $\hat{f}_{crit}$; in the context of kernel estimates, $\alpha$ is the window width $h$, while for maximum penalized likelihood estimates $\alpha = \lambda$. If the bimodality is strong, we expect $\alpha_{crit}$ to be large. The significance value of $\alpha_{crit}$ can be evaluated by generating a large number of random samples from $\hat{f}_{crit}$, of the same size $n$ as the original data set, and determining the proportion $p$ of such samples that yield a unimodal density estimate using the smoothing parameter $\alpha_{crit}$. This proportion is equal to the significance level of the unimodal hypothesis.

Silverman's test is designed purely to uncover multimodality; the width and separation of the peaks is irrelevant except insofar as they affect the degree to which the peaks can be discerned in a given data set.

Table 1 gives $\alpha_{crit}$ and $p$ as determined from estimates of $f$, $N_o$ and $N_p$ made using the three techniques described above. In no case can the unimodal hypothesis be rejected with reasonable certainty, i.e. in no case is $p$ substantially less than one. We conclude that there is no significant evidence in this sample for a bimodal population of apparent or intrinsic shapes.

We will reach a different conclusion about the possible significance of bimodality below, when we discuss the triaxial inverse problem.

## 3. THE TRIAXIAL INVERSE PROBLEM

Here we demonstrate that relaxing the assumption of axisymmetry allows the construction of physically reasonable intrinsic-shape frequency functions that are fully consistent with the data. We present several possible frequency functions $N(\beta_1, \beta_2)$ of galaxies whose principal axes have lengths in the ratio $1 : \beta_1 : \beta_2$ ($1 \geq \beta_1 \geq \beta_2$). Finally, we apply the principle of maximum entropy (Jaynes 1957) to determine the "most probable" of these distributions.

Before doing so, we note one qualitative way in which triaxial galaxies differ from axisymmetric ones in terms of their statistical projection onto the plane of the sky. Figure 7 shows the function $f(q; \beta_1, \beta_2)$ describing the probability that a single triaxial ellipsoid with axis ratios $(\beta_1, \beta_2)$ will appear, under random projection, as an ellipse of apparent axis ratio $q$. (See also Binney & de Vaucouleurs 1981, Fig. 1.) The function $f(q; \beta_1, \beta_2)$ is finite at $q = 1$ for oblate and prolate models, reflecting the fact that axisymmetric galaxies (particularly oblate ones) often appear nearly round in projection. By contrast, for any triaxial galaxy, $f(q; \beta_1, \beta_2)$ falls to zero at $q = 1$, since a triaxial galaxy looks round from only a single orientation. The frequency function of apparent shapes constructed from any reasonable distribution of triaxial intrinsic shapes will therefore also fall to zero near $q = 1$. This property of triaxial ellipsoids provides a natural — though, as discussed above, not unique — explanation for the behavior of $\hat{f}(q)$ near $q = 1$ found from the Djorgovski-Ryden sample.

As has often been remarked, there are likely to be many populations of triaxial intrinsic shapes $N(\beta_1, \beta_2)$ that are equally consistent with any observed $f(q)$. One way of attacking the triaxial inverse problem is to place some *ad hoc* restriction on the form of $N$. We begin by assuming that elliptical galaxies are all triaxial to the "same" degree, i.e. that

$$Z = \frac{1 - \beta_1}{1 - \beta_2} \tag{18}$$

is the same for every galaxy. For a population of oblate galaxies, $Z = 0$ and for prolate galaxies $Z = 1$. Under this hypothesis, the frequency function of intrinsic shapes becomes a univariate function, which may be written $N_Z(\beta_2)$. We expect this function to be uniquely defined given an observed $f(q)$.

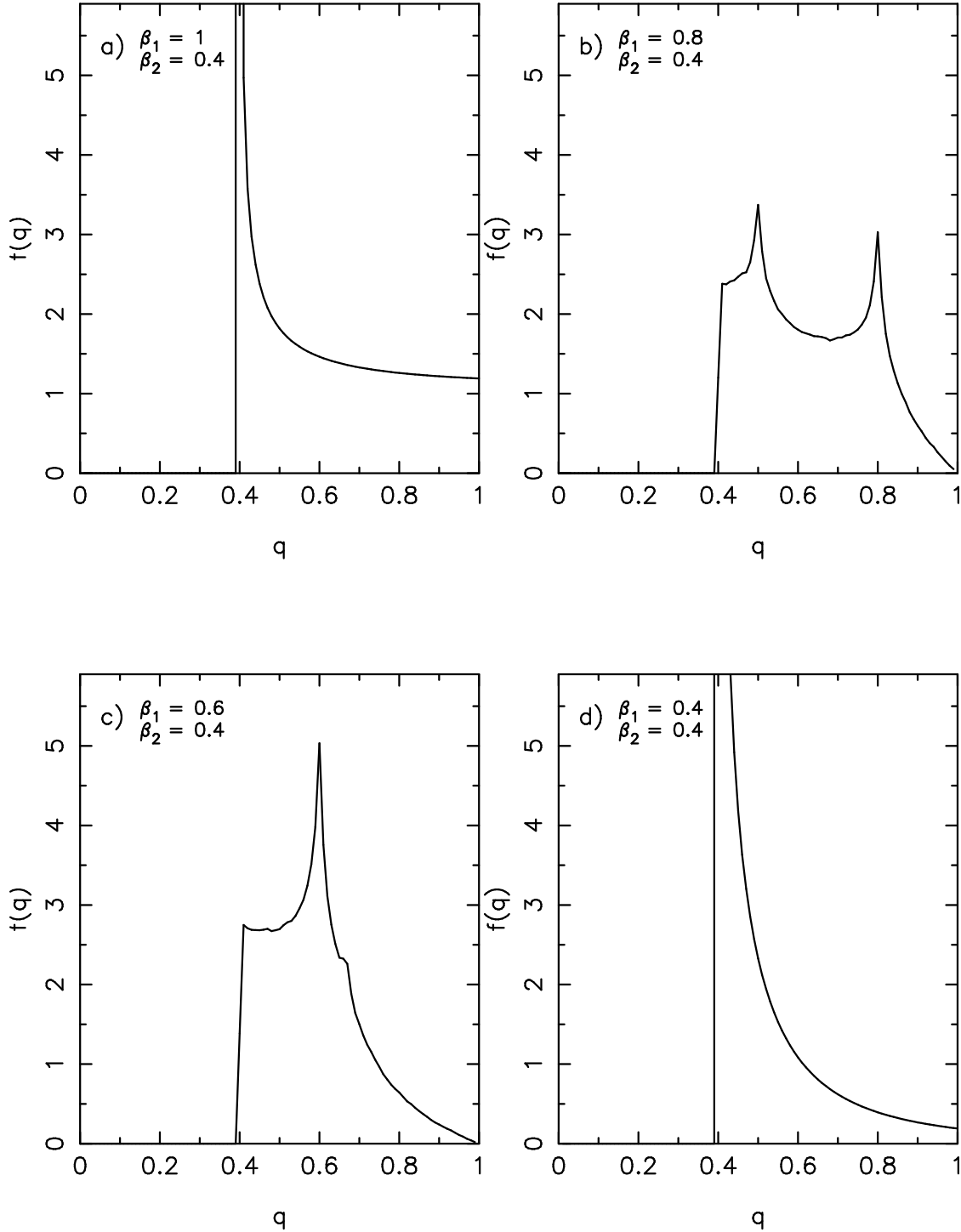

Figure 7. The probability density that a single triaxial ellipsoid whose axis lengths are in the ratio $\beta_1 : \beta_2 : 1$ has an apparent axis ratio $q$.

Under the axisymmetric hypothesis, the relation between $N$ and $f$, and its inverse, could be written explicitly. No such simple relations exist under the triaxial hypothesis. Instead we computed $N_Z(\beta_2)$ according to the following scheme.

1. Construct a smooth estimate of $f(q)$ from the data. We used the maximum penalized likelihood estimate of $f(q)$ defined above, with smoothing parameter $\lambda = 1 \times 10^{-8}$.
2. Define a grid of equal-area cells in $(\beta_1, \beta_2)$ space.

3. From every grid cell, select a random sample of $10^4$ galaxies and view each one from a randomly selected angle. Compute the frequency functions $P_j(q)$ of apparent axis ratios of the galaxies in cell $j$. We used equations (3) in Ryden (1992) to compute the apparent axis ratios; we then used an adaptive kernel algorithm to estimate the smooth functions $P_j$ from the Monte-Carlo data.

4. Find the cell occupation numbers $\hat{N}_j$ that minimize the functional

$$\sum_i \left[ \sum_j N_j P_j(q_i) - f(q_i) \right]^2 + \alpha \int_0^1 \left[ \frac{d^2 N}{d\beta_2^2} \right]^2 d\beta_2, \qquad (19)$$

where the index $i$ denotes the grid in observed axis ratios $q$, and the index $j$ refers to the $(\beta_1, \beta_2)$ grid. As in the penalized likelihood problem discussed above, the first term in (19) measures the deviation of the model from the data, while the second measures the roughness of the model. The role of the smoothing term is slightly different here, however. The source of the instability is now the transformation from $q$ space to $\beta_2$ space, an operation that is equivalent to solving an integral equation for $N$ given $f$. Such equations are often ill-conditioned (e.g. Miller 1974) and the penalty function term is intended to deal with that ill-conditioning. In its absence, the solution $\hat{N}$ would be extremely grid-dependent and unsmooth. We selected the smoothing parameter $\alpha$ by requiring our solutions to reproduce the axisymmetric solutions $N_o$ and $N_p$, derived above by operating directly on $f(q)$, in the axisymmetric cases $Z = 0$ and $Z = 1$.

Figure 8 shows $\hat{N}_Z(\beta_2)$ for several values of the triaxiality index $Z$. The intrinsic frequency function is significantly negative when $Z \lesssim 0.2$, i.e. approximately oblate, and for $Z \gtrsim 0.9$, approximately prolate. The dip below zero always takes place near $\beta_1 \approx \beta_2 = 1$, as in the axisymmetric cases treated above. For $0.3 \lesssim Z \lesssim 0.8$, however, $\hat{N}_Z(\beta_2)$ lies above zero with 90% confidence at all axis ratios. Thus we conclude that the Djorgovski-Ryden data are consistent with a single triaxial population as long as the triaxiality is fairly strong, $0.3 \lesssim Z \lesssim 0.8$.

Interestingly, for $Z \approx 1/2$, the frequency function of intrinsic shapes is very strongly bimodal, with peaks near $\beta_2 = 0.6$ and $\beta_2 = 0.8$. This bimodality is so strong that we felt no need to test its statistical significance via Silverman's "bump-hunting" algorithm described above. Of course, the *physical* significance of this bimodality depends entirely on the reasonableness of our restriction to a single value of $Z$. However it is interesting that such a wide range of frequency functions – from essentially unimodal to strongly bimodal, etc. – can describe the observed distribution of Hubble types equally well. Clearly there is no justification, based on the data alone, for preferring one functional form for $N$ over another.

Given this state of affairs, it is reasonable to search for the "maximally uninformed" solution – that is, the expression for $N$ that has the combinatorially highest probability of occurring given the constraints imposed by the data. Following Jaynes (1957), we therefore seek to maximize the informational entropy:

$$ - \int \int N(\beta_1, \beta_2) \log N(\beta_1, \beta_2) d\beta_1 d\beta_2 $$

subject to the constraint that $N$ generate the observed frequency function of Hubble types. This is equivalent to choosing the $N_j$ to minimize

$$ \sum_i \left[ \sum_j N_j P_j(q_i) - f(q_i) \right]^2 + \delta \sum_j N_j \log N_j \qquad (20) $$

where $\delta$ is chosen large enough that $\hat{N}$ is smooth, but small enough that $\sum_j \hat{N}_j P_j(q)$ is not too different from $f(q)$ (Gull 1989). Note that – since $N(\beta_1, \beta_2)$ is no longer uniquely constrained by the data – the entropy term plays a double role: it guarantees a smooth solution, like the penalty functions described above, but it also selects the (hopefully unique) functional form for $\hat{N}$ that maximizes the entropy.

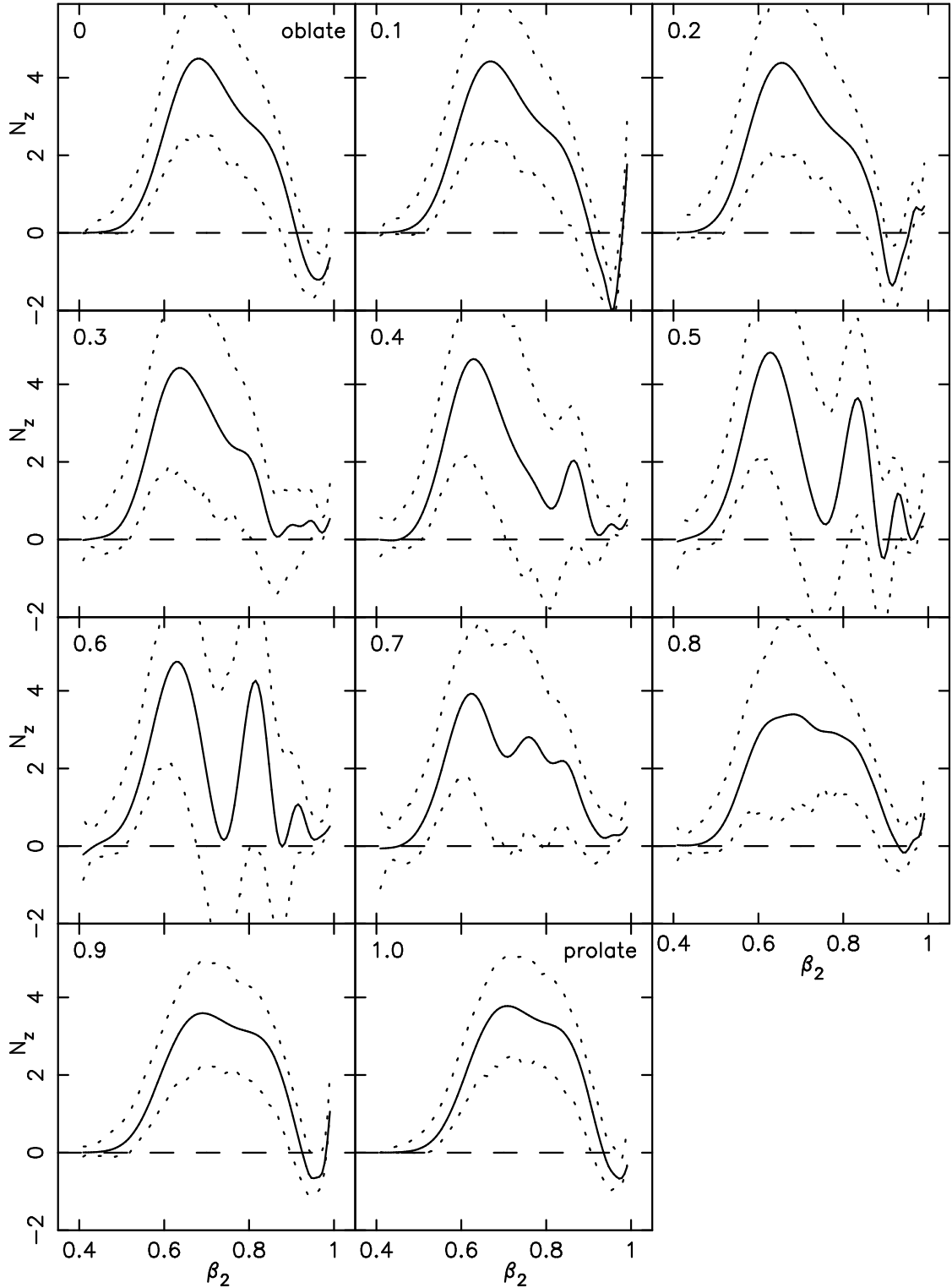

Figure 8. The frequency function of intrinsic shapes under the assumption that all galaxies are triaxial to the "same" degree, i.e. $Z = (1 - b/a)/(1 - c/a) =$ constant. The value of $Z$ is given in the upper left of each panel. Dashed lines are 90% confidence bands on the estimates.

Figure 9 shows the result. The maximum entropy solution for $N(\beta_1, \beta_2)$ is not enormously different from the frequency functions $\hat{N}_o(\beta)$ derived above under the oblate hypothesis. The density is strongly peaked toward $\beta_1 = 1$, i.e. the oblate axis, and exhibits a moderately bimodal structure with peaks at $\beta_2 \approx 0.65$ and $\beta_2 \approx 0.8$. Crudely speaking, the maximum entropy algorithm appears to modify the oblate solution only to the extent of "moving" it slightly off of the oblate axis, in order to avoid introducing too many apparently round galaxies.

Statler (1994) shows a maximum entropy estimate of $N(\beta_1, \beta_2)$ based also on the Djorgovski-Ryden data set. He displays $N$ in the $(\beta_2, T)$ plane, where $T = (1 - \beta_1^2)/(1 - \beta_2^2)$; after suitably transforming the variables, our plot looks reasonably similar to his.

## 4. CONCLUSIONS

1. Nonparametric function estimation techniques are well suited to the frequency-function inverse problem for galaxy intrinsic shapes. These techniques – especially the "maximum penalized likelihood" method – avoid the pitfalls associated with more *ad hoc* approaches such as Richardson-Lucy iteration or parametric model fits.

2. The observed distribution of Hubble types is inconsistent with both the oblate and prolate hypotheses at a high level of significance. The source of the inconsistency is the absence of apparently round elliptical galaxies.

3. Many distributions of triaxial intrinsic shapes are consistent with the data. If all elliptical galaxies are triaxial to the "same" degree, $Z = (1 - b/a)/(1 - c/a) =$ constant, then values of $Z$ in the range $0.3 \lesssim Z \lesssim 0.8$ are allowed.

4. The "maximum entropy" distribution of intrinsic shapes under the triaxial hypothesis may be found. It is weakly bimodal and weighted toward oblate figures.

We thank B. Ryden for helpful discussions and for lending us the data set on which this paper was based. This work was supported by NSF grants AST 90-16515 and 93-18617 and by NASA grant NAG 5-2803.

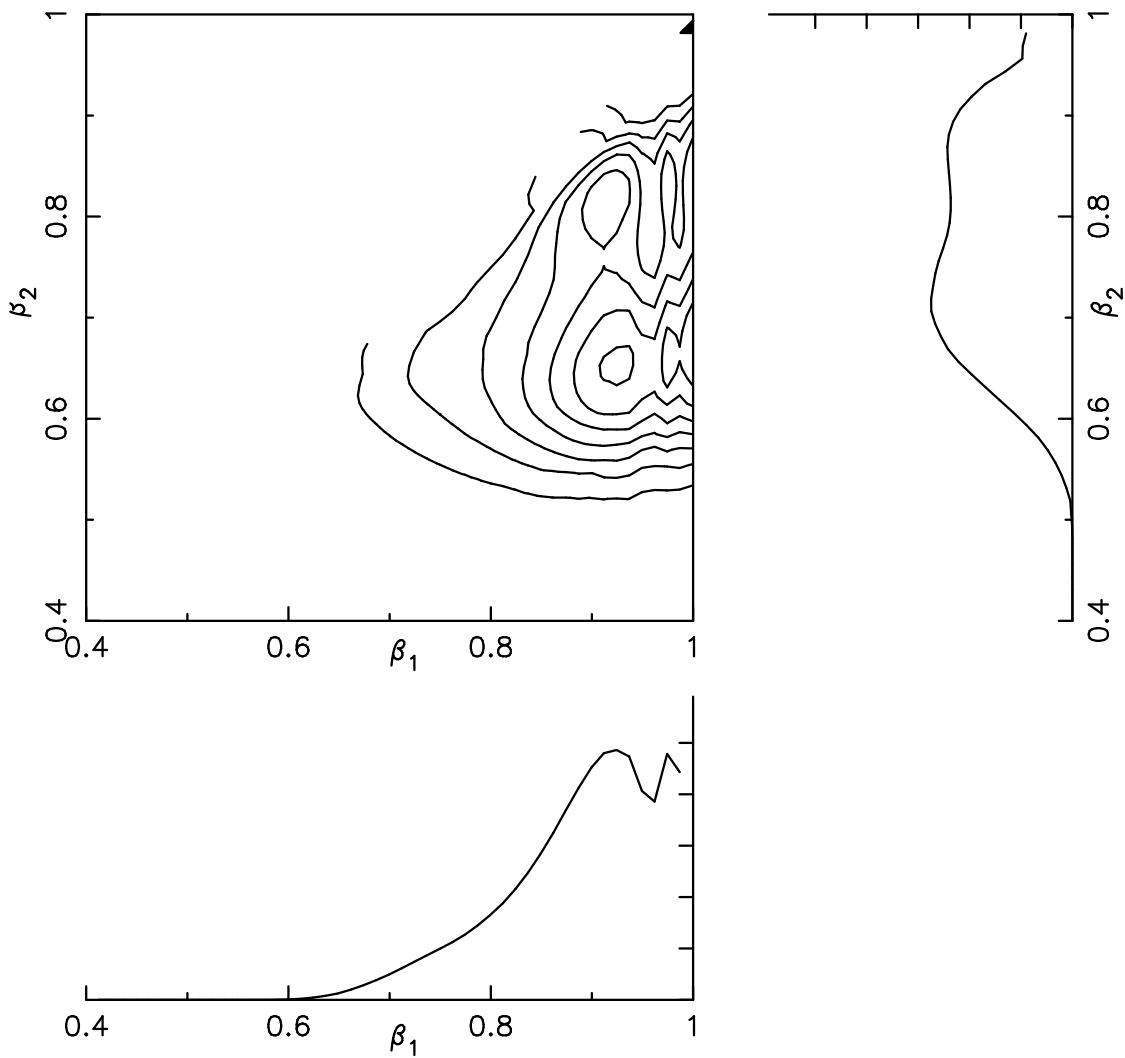

Figure 9. The "maximum entropy" estimate of the frequency function $N(\beta_1, \beta_2)$, normalized to unit total number. Contours are separated by 0.03 in $\hat{N}$. The side plots show the projections of $\hat{N}$ along the two principal axes.

Table 1

Tests of the Bimodal Hypothesis

| Algorithm | Function | $\alpha_{crit}$ | p |
|---|---|---|---|
| MPL | $f(q)$ | $1.14 \times 10^{-9}$ | 0.57 |
| | $N_o(\beta)$ | $3.73 \times 10^{-9}$ | 0.58 |
| | $N_p(\beta)$ | $6.04 \times 10^{-9}$ | 0.67 |
| Kernel[1] | $f(q)$ | 0.10 | 0.50 |
| | $N_o(\beta)$ | 0.11 | 0.54 |
| | $N_p(\beta)$ | 0.115 | 0.38 |
| Kernel[2] | $f(q)$ | 0.56 | 0.71 |
| | $N_o(\beta)$ | 0.69 | 0.84 |
| | $N_p(\beta)$ | 0.71 | 0.81 |

[1] Adaptive kernel with reflection
[2] Adaptive kernel with coordinate transformation

# 5. REFERENCES

Abramson I.S. 1982, J. Mult. Anal., 112, 562
Binney, J. & de Vaucouleurs, G. 1981, MNRAS, 194, 679
Djorgovski, S. 1985, PhD thesis, Univ. California, Berkeley
Efron, B. 1982, The Jackknife, the Boostrap and other Resampling Plans (SIAM: Philadelphia)
Fasano, G. & Vio, R. 1991, MNRAS, 249, 629
Franx, M., Illingworth, G. & de Zeeuw, P. T. 1991, ApJ, 383, 112
Gull, S.F. 1989, in Maximum Entropy and Bayesian Methods, ed. J. Skilling (Kluwer: Dordrecht)
Hall P.G. 1990, J. Mult. Anal. 32, 177
Hubble, E. P. 1926, ApJ, 64, 321
Jaynes, E.T. 1957, Phys. Rev. 106, 620
Lucy, L. B. 1974, AJ, 79, 745
Merritt, D. 1990. Invited Lecture, "Intrinsic Shapes of Elliptical Galaxies," Sant'Agata, Italy, September 5.
Merritt, D. 1992, In "Morphological and Physical Classification of Galaxies," ed. G. Longo, M. Capaccioli & G. Busarello (Kluwer: Dordrecht), p. 309.
Mihalas, D. & Binney, J. J. 1981, Galactic Astronomy (Freeman: San Francisco)
Miller, G.F. 1974, in Numerical Solutions of Integral Equations, L.M. Delves and J. Walsh eds (Oxford:Clarendon)
O'Sullivan, F. 1986, Stat. Sci. 1, 502
Richardson, W. H. 1972, J. Opt. Soc. Am., 62, 55
Rice, J.A. 1986, Journal of Multivariate Analysis, 19, 251
Ryden, B. S. 1992, ApJ, 396, 445
Scott, D. W. 1992, Multivariate Density Estimation (Wiley: New York)
Schweizer F., & Seitzer P. 1992, AJ 104, 1039
Silverman, B. W. 1981, J. Royal Stat. Soc., 43, 97
Silverman, B. W. 1982, Ann. Stat., 10, 795.
Silverman, B. W. 1986, Density Estimation for Statistics and Data Analysis,(Chapman and Hall: London)
Statler, T. 1994. ApJ, 425, 458
Stone M. 1974, J. Royal Stat. Soc. B, 36, 111
Taylor C.C. 1989, Biometrika, 76, 705
Thompson, J. R. & Tapia, R. A. 1990, Nonparametric Function Estimation, Modeling and Simulation (SIAM: Philadelphia)
Wahba, G. 1990, Spline Models for Observational Data (SIAM: Philadelphia)